\documentclass[12pt,subeqn]{article}

\usepackage{amsmath}

\newcommand{\tr}{\operatorname{Tr}}

\newcommand{\EQ}{\begin{equation}}
\newcommand{\EN}{\end{equation}}

\newcommand{\ket}[1]{\left|#1\right\rangle}      

\newcommand{\bear}{\begin{eqnarray}}
\newcommand{\ear}{\end{eqnarray}}
\newcommand{\bt} { \begin{tabular} }
\newcommand{\et}{ \end{tabular} }
\newcommand{\bc} { \begin{center} }
\newcommand{\ec}{ \end{center} }

\newcommand{\btb} { \begin{table} }
\newcommand{\etb}{ \end{table} }

\begin{document}

\topmargin 0pt
\oddsidemargin 5mm
\newcommand{\NP}[1]{Nucl.\ Phys.\ {\bf #1}}
\newcommand{\PL}[1]{Phys.\ Lett.\ {\bf #1}}
\newcommand{\NC}[1]{Nuovo Cimento {\bf #1}}
\newcommand{\CMP}[1]{Comm.\ Math.\ Phys.\ {\bf #1}}
\newcommand{\PR}[1]{Phys.\ Rev.\ {\bf #1}}
\newcommand{\PRL}[1]{Phys.\ Rev.\ Lett.\ {\bf #1}}
\newcommand{\MPL}[1]{Mod.\ Phys.\ Lett.\ {\bf #1}}
\newcommand{\JETP}[1]{Sov.\ Phys.\ JETP {\bf #1}}
\newcommand{\TMP}[1]{Teor.\ Mat.\ Fiz.\ {\bf #1}}

\renewcommand{\thefootnote}{\fnsymbol{footnote}}

\newpage
\setcounter{page}{0}
\begin{titlepage}
\begin{flushright}
UFSCARF-TH-05-22
\end{flushright}
\vspace{0.5cm}
\begin{center}
{\large Integrable Vertex Models with General Twists}\\
\vspace{1cm}
{\large M.J. Martins} \\
\vspace{1cm}
{\em Universidade Federal de S\~ao Carlos\\
Departamento de F\'{\i}sica \\
C.P. 676, 13565-905~~S\~ao Carlos(SP), Brasil}\\
\end{center}
\vspace{0.5cm}

\begin{abstract}
We review recent progress towards the solution of exactly solved isotropic vertex models with
arbitrary toroidal boundary conditions. Quantum space transformations make it possible the diagonalization
of the corresponding transfer matrices by means of the quantum inverse scattering method. Explicit
expressions for the eigenvalues and Bethe ansatz equations of the twisted isotropic spin chains based on the
$B_n$, $D_n$ and $C_n$ Lie algebras are presented. The applicability of this approach to the eight vertex
model with non-diagonal twists is also discussed.
\end{abstract}

\vspace{.15cm}
\vspace{.1cm}
\centerline{Solvable Lattice Models 2004, Kyoto}
\vspace{.15cm}
\centerline{March 2005}

\end{titlepage}


\pagestyle{empty}

\newpage

\pagestyle{plain}
\pagenumbering{arabic}

\renewcommand{\thefootnote}{\arabic{footnote}}

A vertex model is a classical statistical system defined on a lattice whose geometry is given by a possibly infinite
set of straight lines on the plane \cite{BAX}. The intersections of these lines are called vertices or sites
and here we consider the situation where not more than two lines meet at every vertex. Clearly, the square lattice
of size $L \times L$ is the simplest one and from now on we shall restrict ourselves to it. A physical state is then
defined by assigning to each lattice edge a discrete variable having one out of $q$ possible values. 

Next, we suppose
that the corresponding row-to-row transfer matrix can be constructed from elementary local $i$-th site Boltzmann weights
${\cal L}_{{\cal A}i}(\lambda)$ where $\lambda$ denotes a spectral parameter. This operator is best viewed as 
a $ q\times q$ matrix on the auxiliary space $\cal{A} ={\cal C}^{q}$ whose elements are operators acting on the
$\displaystyle{\prod_{i=1}^{L} \otimes {\cal C}_i^{q}}$ Hilbert space. Considering toroidal boundary 
conditions on the square lattice, the transfer matrix $T(\lambda)$ can be written in terms of the trace over
$\cal A$ of the following ordered product of operator \cite{TA,KO,DE} 
\EQ
T(\lambda)=\tr_{\cal A}\left[ 
{\cal G}_{\cal A} {\cal L}_{{\cal A} L}(\lambda){\cal L}_{{\cal A} L-1}(\lambda)\dots {\cal L}_{{\cal A} 1}(\lambda)
 \right]
\label{tra}
\EN
where
${\cal G}_{\cal A}$ 
are $ q\times q$ $c$-number matrices, representing generalized periodic boundary
conditions.

A sufficient condition for integrability, i.e $\left [ T(\lambda),T(\mu) \right ]=0 $ for arbitrary values of
$\lambda$ and $\mu$, is the existence of an invertible matrix 
${\check{R}}(\lambda,\mu)$ satisfying the property \cite{TA,KO}
\EQ
{\check{R}}(\lambda,\mu) {\cal L}_{{\cal A}i}(\lambda) 
\otimes {\cal L}_{{\cal A}i}(\mu) =
{\cal L}_{{\cal A}i}(\mu) \otimes
{\cal L}_{{\cal A}i}(\lambda) 
{\check{R}}(\lambda,\mu) 
\label{alg}
\EN
and  that
the matrix
${\cal G}_{\cal A}$ is a possible representation, without spectral parameter dependence,
of the quadratic algebra (\ref{alg}), namely \cite{DE}
\EQ
\left[ {\check{R}}(\lambda,\mu), {\cal G}_{\cal A} \otimes {\cal G}_{\cal A}\right]=0,
\label{sym}
\EN

As usual the $R$-matrix
${\check{R}}(\lambda,\mu)$ is required to satisfy the famous Yang-Baxter equation
\EQ
{\check{R}}_{23}(\lambda_1,\lambda_2){\check{R}}_{12}(\lambda_1,\lambda_3){\check{R}}_{23}(
\lambda_2,\lambda_3)={\check{R}}_{12}(\lambda_2,\lambda_3){\check{R}}_{23}(\lambda_1,\lambda_3){\check{R}}_{12}(\lambda_1,\lambda_2),
\label{YB}
\EN

In this paper we will consider integrable models whose corresponding $R$-matrices are additive with respect the
spectral parameters, 
${\check{R}}(\lambda,\mu)=
{\check{R}}(\lambda-\mu)$. In this case, the simplest spectral parameter dependent representation of the
Yang-Baxter algebra (\ref{alg}) turns out to be
\EQ
{\cal L}_{{\cal A}i}(\lambda)=P_{{\cal A}i}{\check{R}}(\lambda),
\label{repre}
\EN
where $P_{{\cal A}i}$ is the exchange operator on the space ${\cal A}\otimes {\cal C}^q_{i}$.

If the matrix ${\cal G}_{{\cal A}}$ is non-singular 
it is possible to derive an
integrable quantum spin chain from $T(\lambda)$. Assuming that the operator  
${\cal L}_{{\cal A}i}(\lambda)$ is proportional to the permutator $P_{{\cal A}i}$, say  at 
certain special point $\lambda=0$,  the
corresponding one-dimensional Hamiltonian  reads \cite{BAT},
\EQ
{\cal H}=\sum_{i=1}^{L-1} P_{{\cal A}i}\frac{d{\cal L}_{{\cal A}i}(\lambda)}{d\lambda}\mid_{\lambda=0}+{\cal G}_{L}^{-1}P_{L 1}\frac{d{\cal L}_{L 1}(\lambda)}{d\lambda}\mid_{\lambda=0}{\cal G}_{L}
\label{ham}
\EN

When the boundary matrix 
${\cal G}_{\cal A}$ is non-diagonal, the diagonalization of either the transfer matrix (\ref{tra}) or the
Hamiltonian (\ref{ham}) is indeed a highly non-trivial problem in the field of integrable models. The main
difficulty is concerned with the apparent lack of simple references states to start the Bethe ansatz analysis.
Here we would like to present the steps towards the direction of solving integrable isotropic vertex models
with non-diagonal toroidal boundary conditions. As concrete examples we will consider those
systems whose rational $R$-matrices are invariant by   
the $B_n$, $D_n$ and  $C_n$ symmetries. 
One way to construct rational solutions of the Yang-Baxter equation (\ref{YB}) is by means of the
braid-monoid algebra \cite{BRA}  at its degenerated point \cite{CHI}. 
This 
algebra is generated by the identity $I_i$, by a braid $b_i$ and a 
Temperley Lieb operator $E_i$ acting on sites $i$ of a chain
of length $L$. On the degenerate point the braid operator becomes a generator
of the symmetric group, namely
\EQ
b_i= \sum_{a,b=1}^{q} \hat{e}_{ab}^{(i)} \otimes \hat{e}_{ba}^{(i+1)}
\EN
where $\hat{e}_{ab}^{(i)}$ are the $q \times q$ Weyl matrices acting on the space
${\cal C}_i^q$. 

The monoid turns out to be represented by the following expression \cite{MAR}
\EQ
E_i= \sum_{a,b,c,d=1}^{q} \alpha_{ab} \alpha^{-1}_{cd} \hat{e}_{ac}^{(i)} \otimes \hat{e}_{bd}^{(i+1)}
\EN
where the matrix $\alpha$ for the models 
$B_n$, $D_n$ and $C_n$ are given by
\EQ
\alpha_{B_n} = {\cal{I}}_{2n+1 \times 2n+1 },~~ \alpha_{D_n} =
{\cal{I}}_{ 2n \times 2n },~~  
\alpha_{C_n}=\left( \begin{array}{cc} 

	O_{n \times n} &     {\cal{I}}_{n \times n} \\
	-{\cal{I}}_{n \times n} &   O_{n \times n}  \\
	\end{array}
	\right)
\EN
such that  ${\cal{I}}_{ k \times k} $ is defined as a $ k \times k $ anti-diagonal matrix.

The set of algebraic relations satisfied by the braid and monoid at its generated point can be ``Baxterized'' in terms
of rational functions. More specifically, the solution $\check{R}(\lambda)$ in terms of  combinations
of the identity, braid and monoid is given by \cite{CHI,MAR}
\EQ
\check{R}_{i,i+1}(\lambda)= I_i+\lambda b_i -\frac{\lambda}{\lambda-\delta}E_i
\EN
where the values of parameter $\delta$ are
\EQ
\delta_{B_n}=-n+\frac{1}{2}~~~
\delta_{C_n}=-n-1~~~
\delta_{D_n}=-n+1
\EN

Let us now turn our attention to the diagonalization of the transfer matrix (\ref{tra}) for the 
above $B_n$, $D_n$ and $C_n$ vertex models, \footnote{ We recall that similar problem
for the $A_n$ model has been tackled  in ref.\cite{RMG}.} considering the most general admissible boundary matrix  
satisfying the condition (\ref{sym}), i.e
$\left [ E_i,{\cal G}_{\cal A} \otimes
{\cal G}_{\cal A} \right ]=0 $.  Denoting by  
$M_{\cal A}$ the matrix that diagonalize the boundary matrix $\cal G_{A}$  and by inserting the
terms 
$M_{\cal A} 
M_{\cal A}^{-1}$ all over the trace (\ref{tra}) one 
we can write $T(\lambda)$ as
\bear
T(\lambda)&=&\tr_{\cal A}{\left[ M_{\cal A} D_{\cal A} M_{\cal A}^{-1} M_{\cal A} \left( M_{\cal A}^{-1} {\cal L}_{{\cal A}L}(\lambda) M_{\cal A} \right) \dots \left( M_{\cal A}^{-1} {\cal L}_{{\cal A}1}(\lambda) M_{\cal A} \right) M_{\cal A}^{-1} \right]}, \label{tttransfer} \\
&=&\tr_{\cal A}{\left[ D_{\cal A} \widetilde{\cal L}_{{\cal A}L}(\lambda) \widetilde{\cal L}_{{\cal A}L-1}(\lambda) \dots \widetilde{\cal L}_{{\cal A}1}(\lambda) \right]},
\label{nada}
\ear
where $D_{\cal A}$ is diagonal matrix whose entries are the eigenvalues of $\cal G_{A}$ and the $\widetilde{\cal L}$-operators are given by
\EQ
\widetilde{\cal L}_{{\cal A}i}(\lambda)=M_{\cal A}^{-1} {\cal L}_{{\cal A}i}(\lambda)M_{\cal A}.
\label{aux}
\EN

The next step in our approach is to note that it is always possible to choose an invertible transformation
$U_i$ on the space ${\cal C}_i^q$ such that
\EQ
U_i^{-1}\widetilde{\cal L}_{{\cal A}i}(\lambda)U_i={\cal L}_{{\cal A}i}(\lambda)
\label{aux1}
\EN
and therefore to undo the modifications on the Lax operators (\ref{aux}) by means of quantum space
transformations.

Now one can take advantage of this remarkable   
property by defining a new transfer matrix $T'(\lambda)$
\bear
T'(\lambda)=\prod_{j=1}^{L} \otimes U_{j}^{-1} T(\lambda) \prod_{j=1}^{L} \otimes U_{j} 
=\tr_{\cal A}{\left[  D_{\cal A} {\cal L}_{{\cal A}L}(\lambda)  \dots {\cal L}_{{\cal A}1}(\lambda) \right]},
\label{TlinhaT}
\ear
which is precisely the transfer matrix of the vertex model we have started with diagonal twist $D_{\cal A}$.

Because the  boundary
$D_{\cal A}$ is diagonal the transfer matrix $T'(\lambda)$ can be diagonalized with very little difference
from the standard periodic case \cite{MAR}. Furthermore, the operators $T(\lambda)$ and $T'(\lambda)$ share
the same eigenvalues and if $\ket{\psi'}$  is an eigenstate of $T'(\lambda)$ the corresponding eigenvector
$\ket{\psi}$ of $T(\lambda)$ is then  
$\displaystyle{\prod_{j=1}^{L} \otimes U_{j} \ket{\psi'}}$. Considering that the algebraic framework
to diagonalize $T'(\lambda)$ has already been described in ref.\cite{MAR}, there is no need 
to repeat it here, and in what follows we shall present only the final results concerning the
Bethe ansatz equations and eigenvalues of the related Hamiltonian (\ref{ham}).  The expression for
the latter can be written in a compact form in terms of the underlying Cartan matrix
$C_{ab}$  
and the normalized length 
$\eta_{a}$ 
of the roots.  To each $a$-ath root we associate a set of rapidities $\lambda_j^{(a)}$ that satisfy
the following non-linear coupled equations,
\EQ
\left[
\frac{\lambda^{(a)}_{j} -\frac{\delta_{a,1}}{2\eta_{a}}}{\lambda^{(a)}_{j} +\frac{\delta_{a,1}}{2\eta_{a}}} 
\right]^{L} =
\frac{g_{a}}{g_{a+1}}\prod_{b=1}^{n} \prod_{k=1,\; k \neq j}^{m_{b}}
\frac{\lambda^{(a)}_{j}-\lambda^{(b)}_{k} -\frac{C_{a,b}}{2\eta_{a}}}{\lambda^{(a)}_{j}-\lambda^{(b)}_{k} +\frac{C_{a,b}}{2\eta_{a}}}, ~~ j=1, \dots, m_{a} ;~~ a=1, \dots, n
\EN
where $g_a$ is the $a$-th eigenvalue of the matrix
$D_{\cal A}$.  

Before proceeding it should be remarked that due to the constraint 
$\left [ E_i,{\cal G}_{\cal A} \otimes
{\cal G}_{\cal A} \right ]=0 $ not all the eigenvalues $g_a$ are independent.  It turns out that only
the first $n$ ratios $\frac{g_a}{g_{a+1}}$ are indeed arbitrary.
The eigenvalues $E(L)$ of the Hamiltonian (\ref{ham}) are  parameterized  by the variables
$\lambda_j^{(1)}$ by  
\EQ
E(L) = -\sum_{i=1}^{m_{1}} \frac{1}{[\lambda^{(1)}_{i}]^{2} - 1/4} +L
\EN

We expect that these results extend to all isotropic integrable vertex models invariant by the
discrete representations of Lie algebras as well as to superalgebras. Note that such systems possess
a broader class of possible non-diagonal boundary matrices as compared with their trigonometric
counterparts . This means that isotropic vertex models with the most general twisted boundary conditions
are in fact genuine systems that deserve to be studied independently. They are also of potential physical
interest, since suitable combinations between non-diagonal boundaries $\cal{G}_{\cal A} $ and rational
$\cal{L}$-operators can described  interesting solvable atom-fields models \cite{HI}.

A natural question to be asked is whether or not this approach can also be of utility
for non-rational vertex models. A tantalizing problem would be the solution of the eight-vertex model
in the presence of non-diagonal toroidal boundary conditions.  The symmetrical eigth-vertex model \cite{BAX}
possesses four different Boltzmann weights $a(\lambda)$, $b(\lambda)$, $c(\lambda)$ and $d(\lambda)$ whose
local operators
${\cal{L}}_{{\cal A}i}(\lambda)$ which are given by the following $2 \times 2$
matrix
\begin{equation}
{\cal{L}}_{{\cal A} i}(\lambda)=\left(\begin{array}{cc} 
a(\lambda)\sigma_i^{+}\sigma_i^{-}+b(\lambda) \sigma_i^{-}\sigma_i^{+} & 
d(\lambda)\sigma_i^{+}+c(\lambda)\sigma_i^{-} \\
c(\lambda)\sigma_i^{+}+d(\lambda) \sigma_i^{-} & 
b(\lambda)\sigma_i^{+}\sigma_i^{-}+a(\lambda) \sigma_i^{-}\sigma_i^{+}  \\
\end{array}  \right)
\label{8ver}
\end{equation}
and $\sigma_i^{\pm}$ are Pauli matrices 
acting on the $i$-th sites of
an one-dimensional lattice of size L. 

This vertex model is known to be solvable in the manifold
\EQ
2\Delta=\frac{a^2(\lambda)+b^2(\lambda)-c^2(\lambda)-d^2(\lambda)}{a(\lambda)b(\lambda)+c(\lambda)d(\lambda)}~~~~
\Gamma=\frac{a(\lambda)b(\lambda)-c(\lambda)d(\lambda)}{
a(\lambda)b(\lambda)+c(\lambda)d(\lambda)}
\EN
where $\Delta$ and $\Gamma$ are arbitrary constants. A possible non-diagonal twist compatible with
integrability is given by
\EQ
{\cal G}_{\cal A} = \left ( \begin{array}{cc}
0 & 1 \\
1 & 0 \\
\end{array}  \right)
\label{bou}
\EN

One can now follow the same steps discussed above. Though we could not undo completely  the modifications
carried out on the auxiliary space due to the manipulations (\ref{nada}) 
we find out that the following quantum space transformation
\EQ
\bar{{\cal L}}_{{\cal A}i}(\lambda)
=U_i^{-1}\widetilde{\cal L}_{{\cal A}i}(\lambda)U_i
\EN
where the matrix $U_i$ is given by
\EQ
U_i = \left ( \begin{array}{cc}
1 & -1 \\
1 & 1 \\
\end{array}  \right)
\label{sim}
\EN

This transformation leads us to an
operator $\bar{\cal L}_{{\cal A}i}(\lambda)$ that preserves the eight vertex form (\ref{8ver}),
\begin{equation}
\bar{{\cal{L}}}_{{\cal A} i}(\lambda)=\left(\begin{array}{cc} 
\bar{a}(\lambda)\sigma_i^{+}\sigma_i^{-}+\bar{b}(\lambda) \sigma_i^{-}\sigma_i^{+} & 
\bar{d}(\lambda)\sigma_i^{+}+\bar{c}(\lambda)\sigma_i^{-} \\
\bar{c}(\lambda)\sigma_i^{+}+\bar{d}(\lambda) \sigma_i^{-} & 
\bar{b}(\lambda)\sigma_i^{+}\sigma_i^{-}+\bar{a}(\lambda) \sigma_i^{-}\sigma_i^{+}  \\
\end{array}  \right)
\end{equation}
where the new Boltzmann weights
$\bar{a}(\lambda)$, $\bar{b}(\lambda)$, $\bar{c}(\lambda)$ and $\bar{d}(\lambda)$ are given by
\EQ
\bar{a}(\lambda)=\frac{a(\lambda)+b(\lambda)+c(\lambda)+d(\lambda)}{2}
\EN
\EQ
\bar{b}(\lambda)=\frac{a(\lambda)+b(\lambda)-c(\lambda)-d(\lambda)}{2}
\EN
\EQ
\bar{c}(\lambda)=\frac{a(\lambda)-b(\lambda)+c(\lambda)-d(\lambda)}{2}
\EN
\EQ
\bar{d}(\lambda)=\frac{a(\lambda)-b(\lambda)-c(\lambda)+d(\lambda)}{2}
\label{nul}
\EN
whose invariants are $\bar{\Delta}=\frac{1}{\Delta}$ and $\bar{\Gamma}=\frac{\Gamma}{\Delta}$ \footnote{ This then
reemphasize why the isotropic limit $\Delta=1$ is special under both auxiliary and quantum transformations.}.

As a consequence of that our remaining task now consists in diagonalizing the following transfer matrix,
\EQ
\bar{T}(\lambda)=\tr_{\cal A}\left[ 
{\cal D}_{\cal A} \bar{{\cal L}}_{{\cal A} L}(\lambda)\bar{{\cal L}}_{{\cal A} L-1}(\lambda)\dots 
\bar{{\cal L}}_{{\cal A} 1}(\lambda)
 \right]
\label{ttt}
\EN
where the diagonal boundary matrix is
\EQ
{\cal D}_{\cal A} = \left ( \begin{array}{cc}
1 & 0 \\
0 & -1 \\
\end{array}  \right)
\label{diagB}
\EN

By construction 
$\bar{T}(\lambda)$ (\ref{ttt}) and $T(\lambda)$ given by Eqs.(\ref{tra},\ref{8ver},\ref{bou})
share the same eigenvalues while the eigenvectors are related by the similarity transformation (\ref{sim}).
Though this procedure brings some simplification in the eigenvalue problem, it is not enough
to make the diagonalization of the transfer matrix
$\bar{T}(\lambda)$ amenable by Bethe ansatz analysis. This is because the operator 
$\bar{{\cal{L}}}_{{\cal A} i}(\lambda)$ has no simple local pseudovacuum that annihilate one of its
off-diagonal matrix elements for arbitrary values of the spectral parameter. The standard
way of solving this problem is by means of the so-called Baxter's gauge transformations \cite{BAX,TA}
which unfortunately does not work here since the diagonal boundary 
${\cal D}_{\cal A}$  is not an identity matrix. In other words, the problem of finding gauge
transformations $M_i(\lambda)$ with the conditions that both the transformed operator
$M_{i+1}^{-1}(\lambda)\bar{{\cal{L}}}_{{\cal A} i}(\lambda)M_i(\lambda)$  has a local vacuum 
independent of $\lambda$ and that does not spoil the diagonal property of 
${\cal D}_{\cal A}$  has eluded us so far. An advantage of this approach, however, is that
we can easily identify the existence of at least one case in each the eigenvalue problem
for $\bar{T}(\lambda)$ (\ref{ttt}) can be solved by standard algebraic Bethe ansatz. 
This clearly occurs when the Boltzmann weight $\bar{d}(\lambda)$ (\ref{nul}) is null.  
Direct inspection reveals us that this happens at the point in which the modulus $\cal{K}$ of the
elliptic functions  parameterizing the eight vertex Boltzmann weights becomes unity. More specifically,
at the value ${\cal{K}}=1$ the weights $a(\lambda)$, $b(\lambda)$, $c(\lambda)$ and $d(\lambda)$ are given
by the following expressions
\EQ
a(\lambda)=\tanh[\lambda+\gamma]~~
b(\lambda)=\tanh[\lambda]~~
\EN
\EQ
c(\lambda)=\tanh[\gamma]~~
d(\lambda)=\tanh[\gamma] \tanh[\lambda] \tanh[\lambda+\gamma]
\EN
which due to (\ref{nul}) implies $\bar{d}(\lambda)=0$. 

Thanks to the above simplification it now remains only 
the  diagonalization of a symmetric six vertex model, whose solution has appeared in many different contexts in the
literature.  The result for the eigenvalue $\Lambda(\lambda)$ of $T(\lambda)$ is therefore
\EQ
\Lambda(\lambda)= \left [ \bar{a}(\lambda) \right ]^L\prod_{i=1}^{m} \frac{\bar{a}(\lambda_j-\lambda)}{\bar{b}(\lambda_j-\lambda)}
-\left [ \bar{b}(\lambda) \right ]^L \prod_{i=1}^{m} \frac{\bar{a}(\lambda-\lambda_j)}{\bar{b}(\lambda-\lambda_j)}
\EN
where the weights  $\bar{a}(\lambda)$, $\bar{b}(\lambda)$ and $\bar{c}(\lambda)$ are
\EQ
\bar{a}(\lambda)=\frac{\sinh[\lambda+\gamma]}{\cosh[\lambda]\cosh[\gamma]}~~~
\bar{b}(\lambda)=\frac{\sinh[\lambda]}{\cosh[\lambda+\gamma]\cosh[\gamma]}~~~
\bar{c}(\lambda)=\frac{\sinh[\gamma]}{\cosh[\lambda+\gamma]\cosh[\lambda]}
\EN

The
integers $m \leq L$  parameterize the multiparticle state of $\bar{T}(\lambda)$ and the
Bethe ansatz roots $\lambda_j$ satisfy the equations
\EQ
\left[\frac{\bar{a}(\lambda_j)}{\bar{b}(\lambda_j)} \right ]^{L}= -\prod_{k \neq j}^{m} 
\frac{\bar{b}(\lambda_k-\lambda_j)}{\bar{a}(\lambda_k-\lambda_j)}
\frac{\bar{a}(\lambda_j-\lambda_k)}{\bar{b}(\lambda_j-\lambda_k)},~~~j=1,\dots,m
\EN

It is conceivable that an adaptation of the above ideas might work for the general eight vertex model (\ref{8ver}) 
with the boundary (\ref{bou}). In fact, the case ${\cal{K}}=0$ \cite{BAT} \footnote{In this case, 
the new Boltzmann weights  
$\bar{a}(\lambda)$, $\bar{b}(\lambda)$, $\bar{c}(\lambda)$ and $\bar{d}(\lambda)$ are double periodic as 
compared with 
the original six-vertex weights. This is one way to see why the Bethe ansatz phase-shift  is
expected to be half of that of the six-vertex model with diagonal toroidal conditions \cite{BAT}.}
was solved by means of certain
functional relations even though the eigenvectors structure is not yet known.  If this could be carried out,
even for particular values of the modulus ${\cal{K}}$, it  would be an
important step toward the understanding
of properties of the eight vertex.

\section*{Acknowledgements}
I would like to thank  J. Shiraishi for the invitation to participate in the workshop  
``Solvable Lattice Models 2004'', the Kyoto University for the hospitality and W. Galleas and
G.A.P. Ribeiro for collaborating on ref.\cite{RMG}. This work is partially supported by 
Fapesp (Funda\c c\~ao de Amparo \`a Pesquisa
do Estado de S\~ao Paulo) and
the Brazilian Research Council-CNPq.

\end{document}